\documentclass[11pt,draftcls,onecolumn]{IEEEtran}

\usepackage{cite}
\usepackage[cmex10]{amsmath}
\usepackage{array}
\usepackage{fixltx2e}
\usepackage{amsfonts}
\usepackage{color}
\usepackage{amsmath}
\usepackage{amsthm}
\usepackage{amssymb}
\usepackage{t1enc}
\usepackage{alltt}
\usepackage{epsfig}
\usepackage{mathrsfs}
\usepackage{graphicx,ifthen}
\usepackage{multirow}
\usepackage{citesort}
\usepackage[scaled]{helvet}
\usepackage{booktabs}
\usepackage{wasysym}
\usepackage{arydshln}
\usepackage{enumerate}

\allowdisplaybreaks[1]
\newcommand{\e}{{\rm e}}

\newtheorem{lemma}{Lemma}
\newtheorem{proposition}{Proposition}

\newtheorem{remark}{Remark}
\newtheorem{corollary}{Corollary}
\makeatletter

\newcommand{\Rmnum}[1]{\expandafter\@slowromancap\romannumeral #1@}

\def\bysame{\leavevmode\hbox to3em{\hrulefill}\thinspace}

\begin{document}
\title{On the Outage Capacity of Orthogonal Space-time Block Codes Over Multi-cluster Scattering MIMO Channels}

\author{Lu~Wei,~\IEEEmembership{Member,~IEEE,}
        Zhong~Zheng,~\IEEEmembership{Student~Member,~IEEE,}
        Jukka~Corander,~and~Giorgio~Taricco,~\IEEEmembership{Fellow,~IEEE}%
\thanks{L. Wei and J. Corander are with the Department of Mathematics and Statistics, University of Helsinki, Finland (e-mails: \{lu.wei, jukka.corander\}@helsinki.fi).}%
\thanks{Z. Zheng is with the Department of Communications and Networking, Aalto University, Finland (e-mail: zhong.zheng@aalto.fi).}%
\thanks{G. Taricco is with the Dipartimento di Elettronica, Politecnico di Torino, Italy (e-mail: taricco@polito.it).}%
\thanks{This work will be presented in part at 2014 IEEE International Symposium on Information Theory.}}


\maketitle

\begin{abstract}
Multiple cluster scattering MIMO channel is a useful model for pico-cellular MIMO networks. In this paper, orthogonal space-time block coded transmission over such a channel is considered, where the effective channel equals the product of $n$ complex Gaussian matrices. A simple and accurate closed-form approximation to the channel outage capacity has been derived in this setting. The result is valid for an arbitrary number of clusters $n-1$ of scatterers and an arbitrary antenna configuration. Numerical results are provided to study the relative outage performance between the multi-cluster and the Rayleigh-fading MIMO channels for which $n=1$.
\end{abstract}
\


\section{Introduction}
Deployment of pico-cellular networks in dense teletraffic areas such as train stations, office buildings and airports is becoming increasingly popular due to their abilities to extend coverage areas and increase network capacity. In general, it is difficult to model the pico-cellular channel as it involves a wide range of physical mechanisms. Among these mechanisms, however, a distinctive feature is that the transmitted signal propagates through a sequence of clusters (layers) of scatterers until it reaches the destination. This multi-layered scattering channel is typical in modeling indoor propagation between floors in a building~\cite[Chap.~$13$]{2007Saunders}. For transceivers equipped with multiple antennas, the effective end-to-end channel becomes a product of the multiple input multiple output~(MIMO) channel matrices of each layer. In literature, this multiple cluster scattering MIMO channel was considered in~\cite{2002Muller,2002aMuller}, and physical motivation for this channel model can be found in~\cite[Sec.~$3$]{2002Andersen}.

Despite the needs to understand the fundamental limits, such as the channel capacity, of the multiple scattering MIMO channels, results in this direction are quite limited. Closed-form expressions of the ergodic capacity have been derived respectively in~\cite{2013AKW} and~\cite{2013AIK} for equal and unequal number of scatterers in each cluster. The ergodic capacity scaling law has been established in~\cite{2013WZTH}. However, for practical transmission schemes such as Orthogonal Space-time Block Codes (OSTBCs), the corresponding information-theoretic quantities have not been addressed in literature. OSTBCs are particularly attractive open-loop transmit diversity schemes that decouple the MIMO channel into scalar channels. Thus, decoding is reduced from a vector detection problem to a scalar one, which significantly decreases the decoding complexity~\cite{1999Tarokh,2003Larsson}. Moreover, OSTBCs require little computational cost for encoding and achieve full spatial diversity gain~\cite{2003Larsson}. The use of OSTBCs facilitates the implementation of outer code, i.e. each of the equivalent scalar channels, cf.~(\ref{eq:sigSISO}), can be encoded independently with a powerful outer code such as Turbo code.

We consider OSTBCs coded transmissions over the multiple cluster scattering MIMO channels, and study the corresponding outage capacity. Outage capacity is a relevant performance measure when the transmission of each codeword spans only one or finitely many fading realizations. This is the scenario of a pico-cellular network, where the mobile terminals are moving at walking speed, so that the channel gain, albeit random, varies so slowly that it can be assumed as constant along a coding block~\cite{2001Biglieri}. For such a delay-limited system, the average capacity over the ensemble of channel realizations, i.e. the ergodic capacity, can not characterize the achievable transmission rates~\cite{2001Biglieri}. In this paper, we propose a simple closed-form approximation to the outage capacity of the multiple cluster MIMO channels based on the derived exact moment expressions. The proposed approximation is valid for arbitrary but finite transceiver sizes and scatterers per cluster. The result is obtained by making use of finite-dimensional singular values distribution for products of complex Gaussian matrices as well as the moment based approximation. Interestingly, the proposed approximation becomes exact as the channel degenerates to a conventional Rayleigh fading channel. Simulations are conducted to show the usefulness of the proposed approximation as well as to compare the outage performance with the conventional MIMO channels. Based on the analytical and numerical results, we gain physical insight into the behavior of the outage capacity of the considered channel model.

The rest of the paper is organized as follows. In Section~\ref{sec:model} we outline the system model studied in this paper, which includes the channel model, the signal model as well as the outage capacity formulation. Section~\ref{sec:analysis} is devoted to the analysis of the outage capacity of the considered system model. Simulations are presented in Section~\ref{sec:simu} to examine the outage performance in various realistic scenarios. In Section~\ref{sec:con} we conclude the main findings of this paper. Proofs of all the technical results are provided in the Appendices.

\section{System Model}\label{sec:model}

\subsection{Channel Model}
\begin{figure}[!t]
\centering
\includegraphics[width=4in]{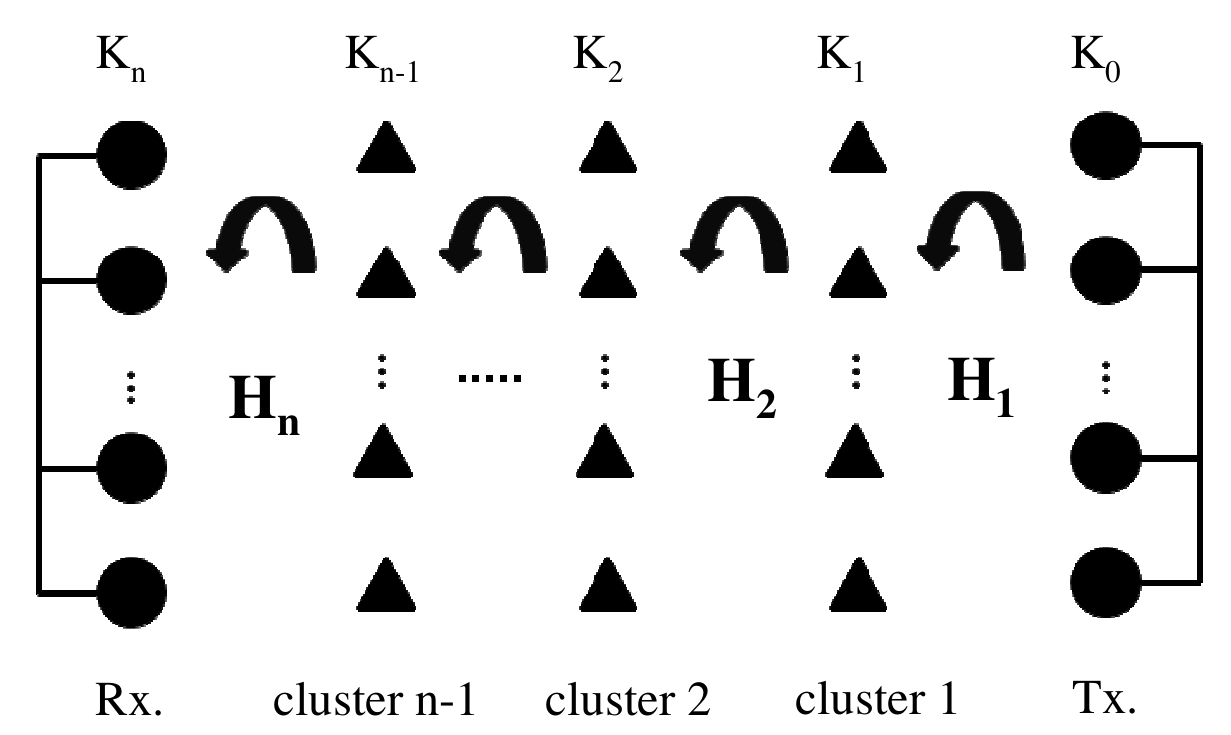}
\caption{Multiple cluster scattering MIMO channel with $n-1$ layers of clusters. The black circle and triangle represent a transmit/receive antenna and a scatterer, respectively.}\label{fig:ch}
\end{figure}

Consider a single user MIMO system with $K_{0}$ transmit and $K_{n}$ receive antennas. Information transmitted to the receiver goes through $n-1$ successive scattering clusters, each having $K_{i}$ ($i=1,\dots,n-1$) scatterers, as shown in Fig.~\ref{fig:ch}. The channels between non-consecutive clusters as well as the direct link between the transmitter and the receiver are ignored. As a result, the effective channel between the transmitter and the receiver equals the product of $n$ channel matrices
\begin{equation}\label{eq:ch}
\mathbf{P}_{n}=\mathbf{H}_{n}\cdots\mathbf{H}_{1},
\end{equation}
where the dimensions of the $i$-th channel $\mathbf{H}_{i}$ are $K_{i}\times K_{i-1}$. Each channel $\mathbf{H}_{i}$ is assumed to be an i.i.d. Rayleigh fading MIMO channel, i.e. the entries of $\mathbf{H}_{i}$ follow the standard complex Gaussian distribution and are independent of each other. The assumption of the i.i.d. Rayleigh channel requires the so-called richly scattered physical environment, where there exist a large number of statistically independent reflected paths with random amplitudes~\cite[Chap.~$7.3.8$]{2005Tse}. Thus, there needs to exist rich scattering environments creating $\mathbf{H}_{i}$ and $\mathbf{H}_{i+1}$. Between these two environments, all scattering happens through the $K_{i}$ scatterers in cluster $i$, which can be thought as $K_{i}$ keyholes. Examples of such channel model include the channel between floors in a building, where inside each floor there is an i.i.d. scattering environment, but between the floors there is restricted propagation through the scatterers~\cite{2007Saunders}. As the number of scatterers of all clusters goes to infinity with the antenna size kept fixed, it is expected that the channel~(\ref{eq:ch}) reduces to a conventional Rayleigh fading channel. This intuitively clear fact will be proven in Section~\ref{sec:rela}. Note that the model~(\ref{eq:ch}) also describes the multi-hop amplify-and-forward MIMO relay channels when assuming noiseless relays~\cite{2011Fawaz}. Obviously, for $n=1$ the channel~(\ref{eq:ch}) becomes the conventional MIMO channel. We notice that the channel~(\ref{eq:ch}) is also referred to as the `Rayleigh product MIMO channel' in literature~\cite{2008Jin}.

\subsection{Signal Model}
We consider linear space-time coded transmissions over the multiple cluster scattering MIMO channels~(\ref{eq:ch}). We assume quasi-static flat fading channels. Namely, the channel remains constant for at least the transmission of an entire frame (say $T$ symbols), and may vary from frame to frame. The resulting signal model within one frame reads
\begin{equation}\label{eq:sig}
\mathbf{Y}=\frac{\mathbf{P}_{n}}{\sqrt{\mathcal{N}}}\mathcal{G}+\mathbf{W},
\end{equation}
where the $K_{n}\times T$ matrix $\mathbf{Y}$ denotes the received signals. The entries of the $K_{n}\times T$ noise matrix $\mathbf{W}$ are i.i.d. and follow the standard complex Gaussian distribution. In line with the convention~\cite{2002Muller,2013AKW,2013AIK}, the effective channel $\mathbf{P}_{n}$ is normalized by $\mathcal{N}=\prod_{i=1}^{n}K_{i}$ so that the total energy of the normalized channel\footnote{$\mathrm{tr}(\cdot)$ denotes the matrix trace operation, and $(\cdot)^{\dag}$ denotes the conjugate-transpose.} $\mathrm{tr}\left(\mathbb{E}\left[\mathbf{P}_{n}\mathbf{P}_{n}^{\dag}\right]/\mathcal{N}\right)=K_{0}$, cf.~(\ref{eq:m1}), will not grow with $n$. In~(\ref{eq:sig}), the $K_{0}\times T$ matrix $\mathcal{G}$ denotes the linear OSTBC mappings of $S$ transmitted symbols in such a way that $\mathcal{G}\mathcal{G}^{\dag}$ is proportional to an identity matrix. Since the encoding matrix $\mathcal{G}$ spans $T$ symbol times to encode $S$ symbols, the code rate equals $R=S/T$, which is also referred to as the delay-optimality of the code. It is shown in~\cite{1999Tarokh} that full rate $R=1$ OSTBCs exist for any number of transmit antennas using any real constellation such as PAM. For any complex constellation such as PSK/QAM, half rate $R=1/2$ OSTBCs exist for any number of transmit antennas, while full rate OSTBC only exists for two transmit antennas, a.k.a. the Alamouti scheme. For specific cases of two, three, and four transmit antennas, rate $R=1$, $R=3/4$, $R=3/4$ OSTBCs for complex constellations are given in~\cite{1999Tarokh}. Without loss of generality, we assume complex constellations in the following discussions.

Due to the orthogonality property of OSTBCs, i.e. $\mathcal{G}\mathcal{G}^{\dag}\propto\mathbf{I}$, the MIMO channel is decoupled into $S$ independent scalar complex AWGN channels after decoding. The resulting equivalent SISO signal model reads~\cite[Th. 7.3]{2003Larsson}
\begin{equation}\label{eq:sigSISO}
y_{i}=\frac{\left\|\mathbf{P}_{n}\right\|^2_{\text{F}}}{R\mathcal{N}}x_{i}+w_{i},~~~i=1,\dots,S,
\end{equation}
where $x_{i}$ denotes the transmitted symbol and the noise $w_{i}$ follows a complex Gaussian distribution with mean zero and variance $\left\|\mathbf{P}_{n}\right\|^2_{\text{F}}/R\mathcal{N}$. We denote by $\gamma$ the total transmit power per symbol time, which equals the transmit SNR. Here, $\left\|\mathbf{P}_{n}\right\|_{\text{F}}=\sqrt{\mathrm{tr}\left(\mathbf{P}_{n}\mathbf{P}_{n}^{\dag}\right)}$ denotes the Frobenius norm, and we define $X=\left\|\mathbf{P}_{n}\right\|^{2}_{\text{F}}$. With the above notations, the effective SNR of the equivalent signal model~(\ref{eq:sigSISO}) at the output of the STBC decoder equals
\begin{equation}\label{eq:SNR}
\text{SNR}=\frac{\gamma}{RK_{0}\mathcal{N}}\left\|\mathbf{P}_{n}\right\|^{2}_{\text{F}}.
\end{equation}

\subsection{Outage Capacity}
Since the OSTBCs decouple the MIMO channel into independent SISO channels, the problem reduces to the study of the corresponding scalar channels. In particular, the capacity of the multi-cluster scattering MIMO channels (in nats/s/Hz) equals $S$ times the capacity of the SISO system~(\ref{eq:sigSISO}), divided by the number of time instants $T$ used for the transmission:
\begin{equation}\label{eq:MI}
C=R\ln\left(1+\frac{\gamma}{RK_{0}\mathcal{N}}\left\|\mathbf{P}_{n}\right\|^{2}_{\text{F}}\right).
\end{equation}
As we are interested in the delay-limited system, where each codeword sees one channel realization, the fundamental limit of such a system is best explained in the capacity versus outage formalism. Namely, for a given rate $z$ the outage probability, i.e. the Cumulative Distribution Function (CDF) of $C$, is obtained as
\begin{equation}\label{eq:op}
P_{\text{out}}=\mathbb{P}(C<z)=F_{X}\left(\frac{RK_{0}\mathcal{N}}{\gamma}\left(\e^{\frac{z}{R}}-1\right)\right),
\end{equation}
where $F_{X}(\cdot)$ denotes the CDF of $X=\left\|\mathbf{P}_{n}\right\|^{2}_{\text{F}}$. The resulting outage capacity for a given outage probability equals
\begin{equation}\label{eq:oc}
C_{\text{out}}=R\ln\left(1+\frac{\gamma}{RK_{0}\mathcal{N}}F_{X}^{-1}(P_{\text{out}})\right),
\end{equation}
where $F_{X}^{-1}(\cdot)$ denotes the inverse function of $F_{X}(\cdot)$. The outage capacity can be understood as the capacity guaranteed for $1-P_{\text{out}}$ of the transmissions. The benefit of using the above performance metrics is manifested by the fact that the outage probability is directly related to the Packet Error Rate (PER) when codewords span only one fading block. Namely, assuming that the transmitted codeword (packet) is decoded successfully if the transmission rate $z$ is less than the capacity $C$ for the given channel realization $\mathbf{P}_{n}$ and declaring a decoding error otherwise, then the outage probability $P_{\text{out}}$ equals the PER. The outage probability is achievable~\cite{2006Prasad} in the sense that for any $\varepsilon>0$, there exists a code of sufficiently large block length for which the PER is upper-bounded by $P_{\text{out}}+\varepsilon$. Thus, outage capacity provides useful insights on the performance of a delay-limited coded system.

\section{Outage Capacity Analysis}\label{sec:analysis}

\subsection{Exact Moments of $\left\|\mathbf{P}_{n}\right\|^{2}_{\text{F}}$}
It is seen from~(\ref{eq:op}) that analyzing the outage capacity requires the distribution of the random variable $\left\|\mathbf{P}_{n}\right\|^{2}_{\text{F}}$. Since the maximal rank of the channel matrix $\mathbf{P}_{n}$ is
\begin{equation}
K_{\min}=\min\left(K_{0},\ldots,K_{n}\right),
\end{equation}
the Hermitian matrix $\mathbf{P}_{n}\mathbf{P}_{n}^{\dag}$ has $K_{\min}$ nonzero eigenvalues, which we denote by $0<\lambda_{K_{\min}}\leq\ldots\leq\lambda_{1}<\infty$. It is shown in~\cite{2013AIK}, cf.~(\ref{eq:ker}) and the subsequent discussions, that the joint density of nonzero eigenvalues of $\mathbf{P}_{n}\mathbf{P}_{n}^{\dag}$ is invariant under any permutation of the matrix dimensions $K_{0},\dots,K_{n}$. Thus, without loss of generality we set $K_{0}=K_{\min}$ and denote
\begin{equation}
\nu_{i}=K_{i}-K_{0},~~~i=0,\ldots,n.
\end{equation}
We can now write the random variable of interest as
\begin{equation}\label{eq:}
X=\left\|\mathbf{P}_{n}\right\|^{2}_{\text{F}}=\mathrm{tr}\left(\mathbf{P}_{n}\mathbf{P}_{n}^{\dag}\right)=\sum_{i=1}^{K_{0}}\lambda_{i},
\end{equation}
where the support of $X$ is $[0,\infty)$.

Although the exact distribution of $X$ seems difficult to obtain, simple yet accurate approximations can be constructed based on the moments of $X$. In Propositions $1-2$ and Corollaries $1-2$ we present closed-form expressions of the integer moments of $X$. Before showing these results, we need the following lemma.
\begin{lemma}\label{l:jpdf}
The joint density of the ordered nonzero eigenvalues of $\mathbf{P}_{n}\mathbf{P}_{n}^{\dag}$, $0<\lambda_{K_{0}}\leq\ldots\leq\lambda_{1}<\infty$, reads
\begin{eqnarray}
p^{(n)}_{\mathbf{\Lambda}}(\mathbf{\Lambda})&=&\frac{1}{c}\det\left(\lambda_{i}^{j-1}\right)\det\left(G_{0,n}^{n,0}\left(\lambda_{i}\left|\begin{array}{c}\{\},\{\}\\ \{\nu_{n},\ldots,\nu_{2},\nu_{1}+j-1\},\{\}\\\end{array}\right.\right)\right)\label{eq:jpdf}\\
&=&\frac{1}{K_{0}!}\det\left(\text{ker}(\lambda_{i},\lambda_{j})\right),
\end{eqnarray}
where the so-called correlation kernel $\text{ker}(\lambda_{i},\lambda_{j})$ is given by
\begin{equation}\label{eq:ker}
\text{ker}(\lambda_{i},\lambda_{j})=\sum_{q=0}^{K_{0}-1}G_{1,n+1}^{1,0}\left(\lambda_{i}\left|\begin{array}{c}\{\},\{q+1\}\\ \{0\},\{-\nu_{n},\ldots,-\nu_{1}\}\\\end{array}\right.\right)G_{1,n+1}^{n,1}\left(\lambda_{j}\left|\begin{array}{c}\{-q\},\{\}\\ \{\nu_{n},\ldots,\nu_{1}\},\{0\}\\\end{array}\right.\right).
\end{equation}
Here, $\det(\cdot)$ denotes the matrix determinant\footnote{As in~(\ref{eq:jpdf}), the dimensions of matrices in the determinants in this paper are $K_{0}\times K_{0}$, i.e. $i,j=1,\dots,K_{0}$ unless otherwise stated.} and the function
\begin{equation}\label{eq:MeijerG}
G_{p,q}^{m,n}\left(x\left|\begin{array}{c}\{a_{1},\ldots,a_{n}\},\{a_{n+1},\ldots,a_{p}\}\\\{b_{1},\ldots,b_{m}\},\{b_{m+1},\ldots,b_{q}\}\\
\end{array}\right.\right)=\frac{1}{2\pi\imath}\int_{\mathcal{L}}{\frac{\prod_{j=1}^m\Gamma\left(b_j+z\right)\prod_{j=1}^n\Gamma\left(1-a_j-z\right)}{\prod_{j=n+1}^p \Gamma\left(a_{j}+z\right)\prod_{j=m+1}^q\Gamma\left(1-b_j-z\right)}}x^{-z}\,\mathrm{d}z
\end{equation}
defines the general form of Meijer's G-function, where the contour $\mathcal{L}$ is chosen in such a way that the poles of $\Gamma(b_j+z)$, $j=1,\dots,m$ are separated from the poles of $\Gamma\left(1-a_j-z\right)$, $j=1,\dots,n$. In~(\ref{eq:jpdf}) the constant
\begin{equation}\label{eq:cons}
c=\prod_{j=1}^{K_{0}}\prod_{i=0}^{n}\Gamma(j+\nu_{i}),
\end{equation}
where $\Gamma(\cdot)$ denotes the Gamma function.
\end{lemma}
The proof of Lemma~\ref{l:jpdf} is in~\cite{2013AIK}. In~(\ref{eq:jpdf}) the determinant
\begin{equation}\label{eq:Van}
\det\left(\lambda_{i}^{j-1}\right)=\prod_{1\leq i<j \leq K_{0}}\left(\lambda_{i}-\lambda_{j}\right)
\end{equation}
is a Vandermonde determinant. Note that the corresponding joint eigenvalue density for the product of square matrices, i.e. $\nu_{n}=\ldots=\nu_{1}=0$ was derived in~\cite{2013AKW}. From the kernel representation~(\ref{eq:ker}) of the joint density, it is seen that the density of non-zero eigenvalues of $\mathbf{P}_{n}\mathbf{P}_{n}^{\dag}$ is invariant under the choice of $K_{\min}$ since the Gamma functions in RHS of~(\ref{eq:MeijerG}) commute. This property is referred to as \emph{weak commutation relations} in~\cite{2013IK}. In our setting it implies that the outage capacity does not depend on the ordering of the clusters of scatterers as long as the signal passes through all the clusters.
\begin{remark}\label{rm:n1}
When $n=1$, by simple residue calculations, the Meijer's G-function in~(\ref{eq:jpdf}) reduces to
\begin{equation}\label{eq:n1Meijer}
G_{0,1}^{1,0}\left(\lambda_{i}\left|\begin{array}{c}\{\},\{\}\\\{\nu_{1}+j-1\},\{\}\\\end{array}\right.\right)=\frac
1{2\pi \imath}\int_{\mathcal{L}}\Gamma(z+\nu_{1}+j-1)\lambda_{i}^{-z}\mathrm{d}z=\e^{-\lambda_{i}}\lambda_{i}^{\nu_{1}+j-1}.
\end{equation}
Consequently, the joint density~(\ref{eq:jpdf}) is simplified to
\begin{eqnarray}
p^{(1)}_{\mathbf{\Lambda}}(\mathbf{\Lambda})&=&\frac{1}{\prod_{j=1}^{K_{0}}\Gamma(j)\Gamma(j+\nu_{1})}\det\left(\lambda_{i}^{j-1}\right)\det\bigg(\e^{-\lambda_{i}}\lambda_{i}^{\nu_{1}+j-1}\bigg)\\
&=&\frac{1}{\prod_{j=1}^{K_{0}}\Gamma(j)\Gamma(j+\nu_{1})}\prod_{1\leq i<j \leq K_{0}}\left(\lambda_{i}-\lambda_{j}\right)^{2}\prod_{i=1}^{K_{0}}\e^{-\lambda_{i}}\lambda_{i}^{\nu_{1}},\label{eq:n1jpdf}
\end{eqnarray}
which, as expected, recovers the eigenvalues density of the complex Wishart distribution, i.e. the conventional MIMO channel.
\end{remark}

Using the joint density in Lemma~\ref{l:jpdf}, we arrive at our first result.
\begin{proposition}\label{p:MGF}
The Moment-Generating Function (MGF) of the random variable $X=\sum_{i=1}^{K_{0}}\lambda_{i}$ equals
\begin{equation}\label{eq:MGF}
M_{X}(s)=\frac{\det\left(\sum_{t=0}^{\infty}\Gamma(i+j+\nu_{1}+t-1)\left(\prod_{q=2}^{n}(j+\nu_{q})_{t}\right)s^{t}/t!\right)}{\prod_{j=1}^{K_{0}}\Gamma(j)\Gamma(j+\nu_{1})},
\end{equation}
where
\begin{equation}\label{eq:Poch}
(a)_{t}=\prod_{i=0}^{t-1}(a+i)=\frac{\Gamma(a+t)}{\Gamma(a)}
\end{equation}
denotes the Pochhammer symbol.
\end{proposition}
The proof of Proposition~\ref{p:MGF} is in Appendix~\ref{a:MGF}.

\begin{remark}\label{rm:MGFn1}
For the conventional MIMO channel $n=1$, the MGF~(\ref{eq:MGF}) is simplified to
\begin{eqnarray}
M_{X}(s)&=&\frac{\det\big(\sum_{t=0}^{\infty}\Gamma(i+j+\nu_{1}+t-1)s^{t}/t!\big)}{\prod_{j=1}^{K_{0}}\Gamma(j)\Gamma(j+\nu_{1})}\nonumber\\
&=&\frac{1}{\prod_{j=1}^{K_{0}}\Gamma(j)\Gamma(j+\nu_{1})}\det\left(\frac{\Gamma(i+j+\nu_{1}-1)}{(1-s)^{i+j+\nu_{1}-1}}\right)\nonumber\\
&=&\frac{1}{\prod_{j=1}^{K_{0}}\Gamma(j)\Gamma(j+\nu_{1})}\frac{\det\big(\Gamma(i+j+\nu_{1}-1)\big)}{(1-s)^{K_{0}K_{1}}}=(1-s)^{-K_{0}K_{1}}\label{eq:eg1},
\end{eqnarray}
where in the last equality we have invoked the identity~\cite[Appx.~18]{Mehta}
\begin{equation}\label{eq:DET}
\det\big(\Gamma(i+j+\nu_{1}-1)\big)=\prod_{j=1}^{K_{0}}\Gamma(j)\Gamma(j+\nu_{1}).
\end{equation}
\end{remark}

By the definition of moment-generating function
\begin{equation}\label{eq:defMGF}
M_{X}(s)=\sum_{i=0}^{\infty}\frac{\mathbb{E}\left[X^{i}\right]}{i!}s^{i},
\end{equation}
the moments $\mathbb{E}\left[X^{i}\right]$ of the random variable $X=\sum_{i=1}^{K_{0}}\lambda_{i}$ can be, in principle, extracted from the MGF~(\ref{eq:MGF}). In particular, by the power series expansion of~(\ref{eq:eg1})
\begin{equation}\label{eq:MGFn1s}
M_{X}(s)=(1-s)^{-K_{0}K_{1}}=\sum_{i=0}^{\infty}\frac{\left(K_{0}K_{1}\right)_{i}}{i!}s^{i},
\end{equation}
the $m$-th moment of $X$ for the case of $n=1$ is identified to be
\begin{equation}\label{eq:mon1}
\mathbb{E}\left[X^{m}\right]=\left(K_{0}K_{1}\right)_{m},
\end{equation}
which is a known result. For an arbitrary $n$, exact representations of $\mathbb{E}\left[X^{m}\right]$ are derived in the following proposition.
\begin{proposition}\label{p:ExMo}
The $m$-th moment of the random variable $X=\sum_{i=1}^{K_{0}}\lambda_{i}$ admits the following representations
\begin{eqnarray}
\mathbb{E}\left[X^{m}\right]&\!\!\!=\!\!\!&\frac{m!\sum_{L}\det\left(\Gamma(i+j+\nu_{1}+a_{j}-1)\left(\prod_{q=2}^{n}(j+\nu_{q})_{a_{j}}\right)/a_{j}!\right)}{\prod_{j=1}^{K_{0}}\Gamma(j)\Gamma(j+\nu_{1})}\label{eq:ExMo}\\
&\!\!\!=\!\!\!&\frac{m!}{\prod_{j=1}^{K_{0}}\Gamma(j)\Gamma(j+\nu_{1})}\sum_{L}\prod_{1\leq i<j \leq K_{0}}\left(a_{i}-a_{j}+i-j\right) \prod_{j=1}^{K_{0}}\frac{\prod_{i=1}^{n}\Gamma(j+a_{j}+\nu_{i})}{\Gamma(a_{j}+1)\prod_{i=2}^{n}\Gamma(j+\nu_{i})},
\end{eqnarray}
where the sum $\sum_{L}$ is over the partitions of $a_{1}+\cdots+a_{K_{0}}=m$ with $a_{i}\in\{0,\dots,m\}$ for $i=1,\ldots,K_{0}$.
\end{proposition}
The proof of Proposition~\ref{p:ExMo} is in Appendix~\ref{a:ExMo}. Note that the above sum over partition can be implemented as
$\sum_{a_{1}=0}^{m}\sum_{a_{2}=0}^{m-a_{1}}\cdots\sum_{a_{K_{0}-1}=0}^{m-a_{1}-\cdots-a_{K_{0}-2}}$ with $a_{K_{0}}$ replaced by $m-\sum_{i=1}^{K_{0}-1}a_{i}$ in the summand.

Explicit expressions for the first three moments and the leading order term of higher moments can be derived based on Propositions~\ref{p:MGF} and~\ref{p:ExMo}. Before presenting these results, we need the following matrix determinant identity.
\begin{lemma}\label{l:Det}
The determinant
\begin{equation}\label{eq:DetL2}
\begin{array}{c:c}
\det\Bigg(\Gamma\big(i+j+\nu_{1}-1\big)&\Gamma\big(i+K_{0}+\nu_{1}+m-1\big)\Bigg),
\end{array}
\end{equation}
where the size of the matrix $\big(\Gamma(i+j+\nu_{1}-1)\big)$ is $K_{0}\times(K_{0}-1)$ and $\big(\Gamma(i+K_{0}+m-1)\big)$ is a vector of size $K_{0}\times1$, equals
\begin{equation}
\frac{\Gamma(m+\nu_{1}+K_{0})\Gamma(m+K_{0})}{\Gamma(m+1)}\prod_{i=1}^{K_{0}-1}\Gamma(i)\Gamma(i+\nu_{1}).
\end{equation}
\end{lemma}
The proof of Lemma~\ref{l:Det} is in Appendix~\ref{a:Det}. Note that Lemma~\ref{l:Det} generalizes a known result for which $m=0$ in~\cite[Appx.~18]{Mehta}.

\begin{corollary}\label{c:mo}
The first three exact moments of the random variable $X=\sum_{i=1}^{K_{0}}\lambda_{i}$ are given by
\begin{equation}\label{eq:m1}
\mathbb{E}\left[X\right]=\prod_{i=0}^{n}K_{i}
\end{equation}
\begin{equation}\label{eq:m2}
\mathbb{E}\left[X^{2}\right]=\frac{\prod_{i=0}^{n}K_{i}}{2}\left(\prod_{i=0}^{n}(K_{i}+1)+\prod_{i=0}^{n}(K_{i}-1)\right)
\end{equation}
and
\begin{equation}\label{eq:m3}
\mathbb{E}\left[X^{3}\right]=\frac{\prod_{i=0}^{n}K_{i}}{6}\left(\prod_{i=0}^{n}(K_{i}+2)(K_{i}+1)+4\prod_{i=0}^{n}(K_{i}+1)(K_{i}-1)+\prod_{i=0}^{n}(K_{i}-1)(K_{i}-2)\right),
\end{equation}
respectively.
\end{corollary}
The proof of Corollary~\ref{c:mo} is in Appendix~\ref{a:mo}. Although the first moment~(\ref{eq:m1}) can be also derived via a much simpler probabilistic argument~\cite[Appx.~B]{2013AKW}, this approach can not be applied to obtain~(\ref{eq:m2}) and~(\ref{eq:m3}).

Despite the fact that the exact higher moments $\mathbb{E}\left[X^{m}\right]$, $m>3$, are highly non-trivial to obtain, the leading order term for large $n$ can be identified.
\begin{corollary}\label{c:amo}
The dominant term of higher moments $\mathbb{E}\left[X^{m}\right]$, $m>3$, equals
\begin{equation}\label{eq:mApp}
\mathbb{E}\left[X^{m}\right]\approx\frac{\prod_{i=0}^{n}\left(K_{i}\right)_{m}}{m!}.
\end{equation}
The above term dominates in the sense that
\begin{equation}\label{eq:conApp}
\lim_{n\to\infty}\frac{\mathbb{E}\left[X^{m}\right]}{\prod_{i=0}^{n}\left(K_{i}\right)_{m}/m!}=1,
\end{equation}
which holds for any positive integer $m$.
\end{corollary}
The proof of Corollary~\ref{c:amo} is in Appendix~\ref{a:amo}. In practice, the leading order term~(\ref{eq:mApp}) can be used as an approximation in scenarios when the number of clusters is known/expected to be large.

\subsection{Moment Based Approximation}
Using the derived moment expressions, closed-form approximations to the outage probability can now be constructed. Moment based approximation is a useful tool in situations when the exact distribution is intractable but the analytical moments are available. This is the situation in our case. The basic idea of moment based approximation is to match the moments and support of an unknown distribution by an elementary distribution and the associated orthogonal polynomials~\cite{2006Ha,2009Provost}. Based on this idea, the Gamma distribution and the associated Laguerre polynomials are chosen as $X$ has the same support as the Gamma density. The resulting approximation by matching the first $q$ moments of $X$ can be read off from~\cite[Eq.~(2.7.27)]{2006Ha} as
\begin{equation}\label{eq:OPapp}
F_{X}(x)\approx\frac{\gamma\left(\alpha,x/\beta\right)}{\Gamma(\alpha)}+\epsilon(x),
\end{equation}
where
\begin{equation}
\epsilon(x)=\sum_{i=3}^{q}w_{i}\sum_{j=0}^{i}\frac{(-1)^{j}\Gamma(\alpha+i)}{(i-j)!j!}\frac{\gamma\left(\alpha+j,x/\beta\right)}{\Gamma(\alpha+j)}
\end{equation}
with
\begin{equation}\label{eq:w}
w_{i}=\sum_{l=0}^{i}(-1)^{l}\frac{\Gamma(i+1)\mathbb{E}\left[X^{l}\right]}{(i-l)!l!\Gamma(\alpha+l)\beta^{l}}
\end{equation}
and
$\gamma(a,b)=\int_{0}^{b}t^{a-1}\e^{-t}\mathrm{d}t$ denotes the lower incomplete Gamma function. The parameters
\begin{equation}\label{eq:ab}
\alpha=\frac{\mathbb{E}^{2}\left[X\right]}{\mathbb{E}\left[X^{2}\right]-\mathbb{E}^{2}\left[X\right]},~~~\beta=\frac{\mathbb{E}\left[X^{2}\right]-\mathbb{E}^{2}\left[X\right]}{\mathbb{E}\left[X\right]}
\end{equation}
are calculated by matching the first two moments of $X$ to a Gamma random variable with density
\begin{equation}\label{eq:Ga}
p(x|\alpha,\beta)=\frac{1}{\Gamma(\alpha)\beta^{\alpha}}x^{\alpha-1}\e^{-\frac{x}{\beta}},~~~x\in[0,\infty).
\end{equation}
Thus, the term $\gamma\left(\alpha,x/\beta\right)/\Gamma(\alpha)$ in~(\ref{eq:OPapp}) corresponds to the simplest form of the moment based approximation, where only the first two moments are involved. Inserting~(\ref{eq:OPapp}) into~(\ref{eq:op}) the outage probability is obtained.

\begin{remark}
For the conventional MIMO channel $n=1$, the parameters~(\ref{eq:ab}) reduce to
\begin{equation}\label{eq:n1pa}
\alpha=K_{0}K_{1},~~~\beta=1.
\end{equation}
On the other hand, applying the inverse Laplace transform on the MGF~(\ref{eq:eg1}) the exact distribution of $X$ for $n=1$ is obtained as
\begin{equation}\label{eq:n1df}
\frac{1}{\Gamma(K_{0}K_{1})}\int_{0}^{x}t^{K_{0}K_{1}-1}\e^{-t}\mathrm{d}t=\frac{\gamma\left(K_{0}K_{1},x\right)}{\Gamma(K_{0}K_{1})}.
\end{equation}
Inserting~(\ref{eq:n1pa}) into the proposed approximation~(\ref{eq:OPapp}) and comparing it with~(\ref{eq:n1df}), we observe that when $n=1$ the approximation~(\ref{eq:OPapp}) becomes exact, i.e.
\begin{equation}
F_{X}(x)=\frac{\gamma\left(K_{0}K_{1},x\right)}{\Gamma(K_{0}K_{1})},~~\text{with}~~\epsilon(x)\equiv0.
\end{equation}
\end{remark}
The above fact again justifies the choice of expanding the distribution of $X$ by the Gamma distribution and its associated Laguerre polynomials.

\subsection{Relation to the Outage Capacity of the Conventional MIMO Channel}\label{sec:rela}
As the number of scatterers in each cluster increases, the multi-path richness of the channel increases as well. The following proposition shows that in the limit of infinite scatterers per cluster, the multi-cluster MIMO channel converges to the Rayleigh limit, i.e. the conventional MIMO channel. In particular, this implies the convergence of the outage capacity to the conventional MIMO channel. The proof of Proposition~\ref{p:conv} essentially follows the idea of~\cite{2011Levin}.
\begin{proposition}\label{p:conv}
Define
\begin{equation}\label{ch}
\mathbf{H}=\frac{\mathbf{P}_{n}}{\sqrt{\prod_{i=1}^{n-1}K_i}}
\end{equation}
and $K'=\min\left(K_1,\ldots,K_{n-1}\right)$, in the limit $K_{i}$, $i=1,\dots,n-1$, go to infinity with fixed $\rho_{i}=K_{i}/K'$, $i=1,\dots,n-1$, and fixed antenna size $K_{0}$, $K_{n}$ (this limit is denoted by $K'\rightarrow\infty$ in short), we have
\begin{enumerate}[a)]
\item $\mathbf{H}$ converges in distribution to a standard complex Gaussian random matrix with i.i.d. entries.
\item Let $\Delta_{K'}=\mbox{sup}_{\mathbf{z}}|F_{K'}(\mathbf{z})-\Phi(\mathbf{z})|$, where $F_{K'}(\mathbf{z})$ denotes the joint CDF of $\mathbf{z}=vec(\mathbf{H})$, and $\Phi(\mathbf{z})$ denotes the joint CDF of a standard Gaussian vector. Then $\Delta_{K'}\rightarrow0$ as $K'\rightarrow\infty$, i.e. $F_{K'}(\mathbf{z})$ converges to $\Phi(\mathbf{z})$ uniformly, with at least the same rate as $(K')^{-1/2}\rightarrow0$.
\end{enumerate}
\end{proposition}
The proof of Proposition~\ref{p:conv} is in Appendix~\ref{a:conv}. Note that $vec(\mathbf{H})$ denotes the vector formed by stacking the columns of $\mathbf{H}$. As a direct consequence of Proposition~\ref{p:conv}, the capacity of the multi-cluster MIMO channel~(\ref{eq:MI}) converges in distribution to the conventional MIMO channel as the number of scatterers of all clusters goes to infinity.

\section{Numerical Results}\label{sec:simu}
In this section, we study the outage behavior of the multi-cluster MIMO channels through Monte-Carlo simulations. In particular, we examine the impact of the number of scatterers, cluster sizes as well as OSTBCs with different rates on the outage capacity. In each case, the outage performance of the conventional MIMO channel $n=1$ is included for comparison. Each simulation curve is obtained by averaging over $10^{6}$ independent channel realizations.

\begin{figure}[!t]
\centering
\includegraphics[width=5in]{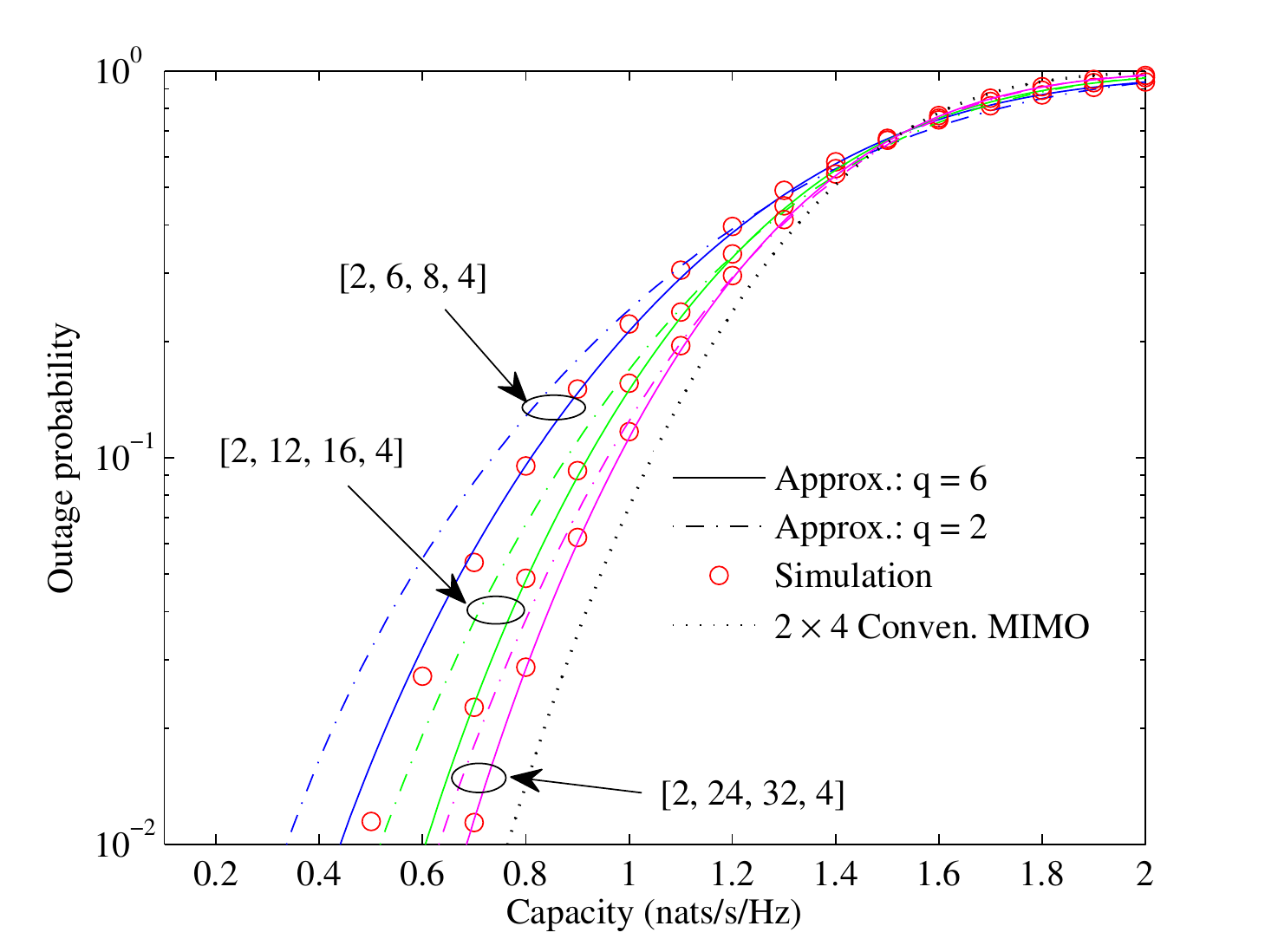}
\caption{Outage probability as a function of capacity: two-clusters MIMO channels with the parameters $[2, K_{1}, K_{2}, 4]$ and its limit of $2\times 4$ Rayleigh-fading MIMO channels.}\label{fig:Pout1}
\end{figure}

\begin{figure}[!t]
\centering
\includegraphics[width=5in]{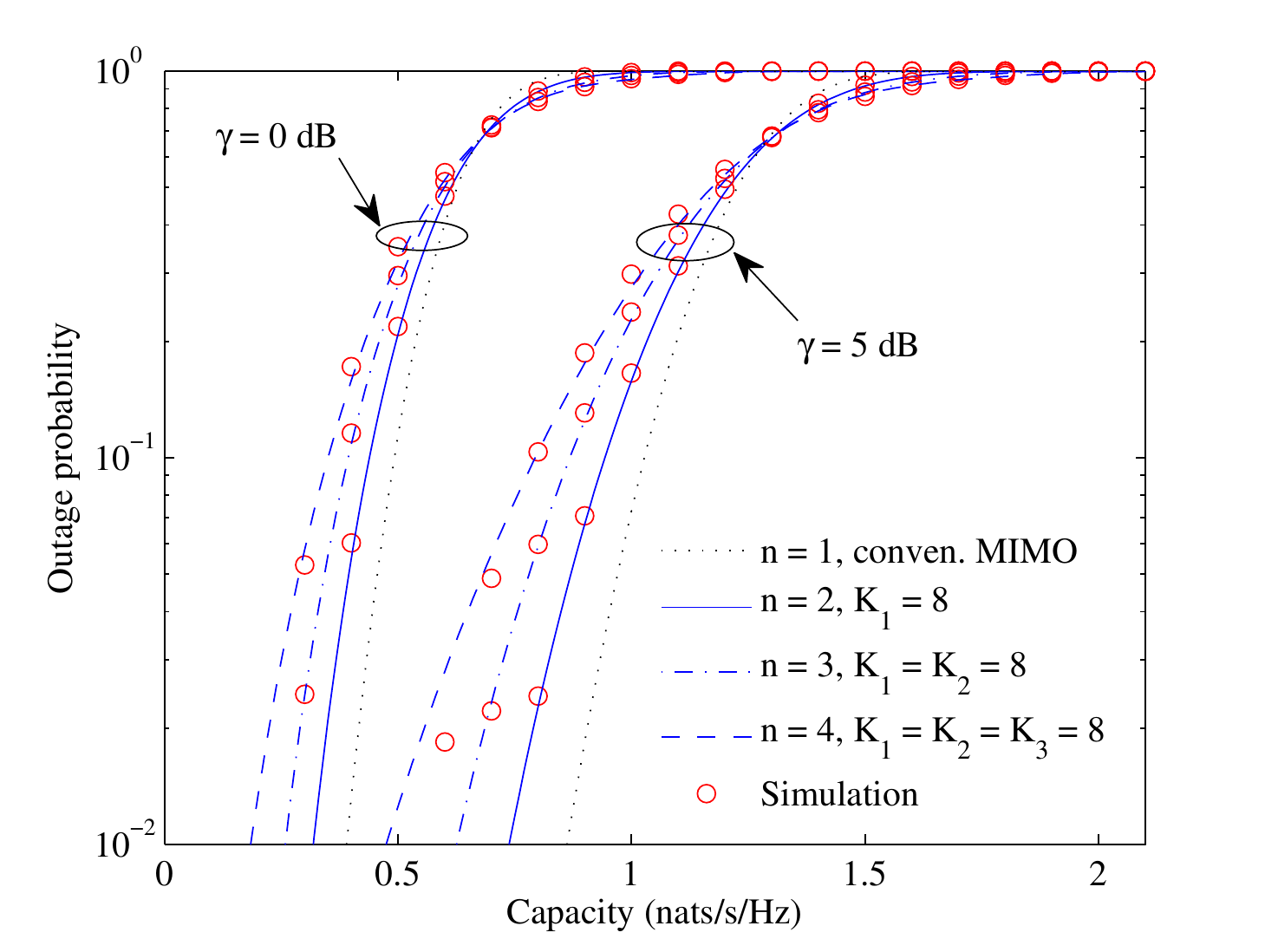}
\caption{Outage probability as a function of capacity: transceiver antenna size $K_{0}=K_{n}=4$ with various number of clusters $n-1$.}\label{fig:Pout2}
\end{figure}

\begin{figure}[!t]
\centering
\includegraphics[width=5in]{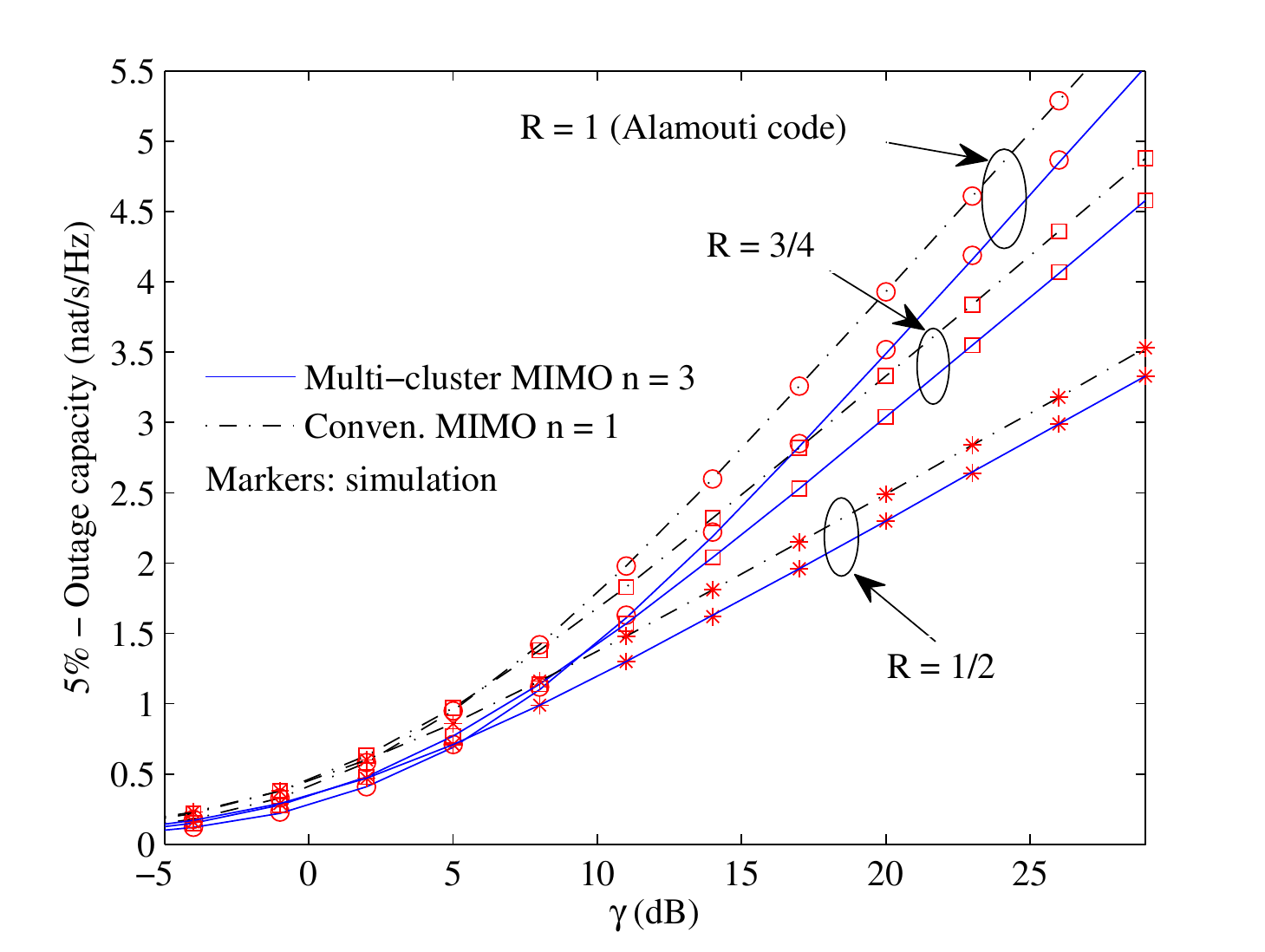}
\caption{Outage capacity as a function of transmit SNR: two-clusters MIMO channels with the parameters $K_{1}=7$, $K_{2}=8$, $K_{3}=4$ and various number of transmit antennas $K_{0}$ with different rates $R$. }\label{fig:Cout}
\end{figure}

In Fig.~\ref{fig:Pout1} the impact of the number of scatterers $K_{i}$, $i=1,\dots,n-1$, and the number of moments $q$ used in the approximation~(\ref{eq:OPapp}) is studied. A scenario of two-cluster scattering MIMO channels, i.e. $n=3$, is considered, where the outage probability~(\ref{eq:op}) is plotted as a function of capacity in nats/s/Hz. The transceiver size is chosen to be $K_{0}=2$, $K_{3}=4$, as such the full rate $R=1$ OSTBC, i.e. Alamouti code, can be used. Various number of scatterers $K_{1}$, $K_{2}$ with a fixed ratio $\rho_{2}=K_{2}/K_{1}=4/3$ is considered. We see from Fig.~\ref{fig:Pout1} that the outage capacity of the multi-cluster MIMO channel is lower than the conventional MIMO channel, which, as predicted by Proposition~\ref{p:conv}, corresponds to the limiting case where the number of the scatterers $K_{1}$, $K_{2}$ goes to infinity. We also observe that as the number of moments~(\ref{eq:ExMo}) increases from $q=2$ to $q=6$, the accuracy of the proposed approximation~(\ref{eq:OPapp}) also increases, as expected. Having seen the effect of $q$ on the approximation accuracy, we set $q=6$ in the remaining figures in order to focus on the impact of other parameters.

In Fig.~\ref{fig:Pout2} we examine the impact of the number of clusters $n-1$ on the outage behavior, where the outage probability~(\ref{eq:op}) is plotted as a function of capacity in nats/s/Hz. The transceiver antenna size is $K_{0}=K_{n}=4$, thus the rate $R=3/4$ OSTBC in~\cite[Eq.~(40)]{1999Tarokh} is used. We consider different number of clusters from $n-1=0$ to $n-1=3$ with an equal scatterer size in each cluster $K_{i}=8$, $i=1,\dots,n-1$. For each $n$, the cases of transmit SNR $\gamma=0$~dB and $\gamma=5$~dB are illustrated. Fig.~\ref{fig:Pout2} shows that increasing the number of clusters leads to a degradation in the outage capacity. This is expected as the presence of a cluster with finite scatterers decreases the multi-path richness of the channel. Moreover, we see that as the outage probability decreases the capacity gap between the multi-cluster and conventional MIMO channels becomes larger. It is also observed that for different number of clusters $n-1$ the outage capacity curves will cross each other, and the one with a larger $n$ achieves a higher outage probability before the crossing and vice versa after the crossing. This phenomenon can be understood by examining the behavior of the sequence of random variables $Y_{n}=\left\|\mathbf{P}_{n}\right\|^2_{\text{F}}/K_{0}\mathcal{N}$ in the capacity expression~(\ref{eq:MI}). Specifically, it can be verified that the sequence of random variables is of the same mean
\begin{equation}
\mathbb{E}\left[Y_{n}\right]=1,~~~\forall n,
\end{equation}
and the variance $\mathbb{V}\left[Y_{n}\right]$ is monotonically increasing with $n$, i.e.
\begin{equation}
\mathbb{V}\left[Y_{n}\right]-\mathbb{V}\left[Y_{n-1}\right]=\frac{1}{2K_{n}}\left(\prod_{i=0}^{n-1}\left(1+\frac{1}{K_{i}}\right)-\prod_{i=0}^{n-1}\left(1-\frac{1}{K_{i}}\right)\right)>0,~~~\forall n.
\end{equation}
For such a sequence of random variables, the CDF plots will intersect each other for different values of $n$. Moreover, for any $n_{1}>n_{2}$ we will have $\mathbb{P}(Y_{n_{1}}<y)>\mathbb{P}(Y_{n_{2}}<y)$ for $y$ before the intersection and vice versa after the intersection.

In Fig.~\ref{fig:Cout} we study the impact of code rate $R$ on the outage performance, where the outage capacity~(\ref{eq:oc}) in nats/s/Hz is plotted as a function of the transmit SNR in dB. The outage probability is set at $P_{\text{out}}=5\%$. A scenario of two-cluster scattering MIMO channels with the number of scatterers $K_{1}=7$ and $K_{2}=8$ is considered. Three different OSTBCs with rates $R=1$ (Alamouti code), $R=3/4$ and $R=1/2$ are considered, where the number of transmit antennas is $K_{0}=2$, $K_{0}=4$ and $K_{0}=8$, respectively. In all cases, the number of receive antennas equals $K_{3}=4$. It is seen from Fig.~\ref{fig:Cout} that increasing the number of transmit antennas does not lead to an improvement of the outage capacity in the high SNR regime. Since in this regime the slope of the outage capacity curve is determined by the code rate $R$, which decreases as the number of transmit antennas increases~\cite{1999Tarokh}. The same phenomenon is also observed for the conventional MIMO channels. Finally, we see from Fig.~\ref{fig:Pout1} to Fig.~\ref{fig:Cout} that the proposed approximation~(\ref{eq:OPapp}) is already reasonably accurate with $q=6$ in all the considered cases.

\section{Conclusion}\label{sec:con}
We study the outage capacity of OSTBCs over the multiple cluster scattering MIMO channels, which is a useful channel model for pico-cellular networks. In such a setting, we derived a simple yet accurate approximation to the outage capacity based on the exact moment expressions. In addition, the relation between the multi-cluster and the conventional MIMO channels has been established. Extensive simulations were conducted to study their relative outage behavior as well as to examine the accuracy of the proposed approximation. Even though the relative outage performance is found to be the same as the conventional MIMO channels, the multi-cluster MIMO channels attain lower outage capacity for systems with a realistic outage rate requirement.

\appendices
\section{Proof of Proposition~\ref{p:MGF}}\label{a:MGF}
Before proving Proposition~\ref{p:MGF}, we need the following matrix integral.

\subsection*{Andr\'{e}ief Integral~\cite{2003Chiani}}
For two $K\times K$ matrices $\mathbf{A}(\mathbf{x})$ and $\mathbf{B}(\mathbf{x})$, with the respective $ij$-th entry being $A_{i}(x_{j})$ and $B_{i}(x_{j})$, and a function $f(\cdot)$ such that $\int_{0}^{\infty}A_{i}(x)B_{j}(x)f(x)\mathrm{d}x<\infty$, the following multiple integral can be evaluated as
\begin{equation}\label{eq:AI}
\int_{\mathcal{D}}\det\big(\mathbf{A}(\mathbf{x})\big)\det\big(\mathbf{B}(\mathbf{x})\big)\prod_{i=1}^{K}f(x_{i})\mathrm{d}\mathbf{x}=\det\left(\int_{0}^{\infty}A_{i}(x)B_{j}(x)f(x)\mathrm{d}x\right),
\end{equation}
where $\mathcal{D}=\{0\leq x_{K}\leq\ldots\leq x_{1}<\infty\}$.

We now start the proof of Proposition~\ref{p:MGF}.
\begin{IEEEproof}
By definition, the MGF of $X=\sum_{i=1}^{K_{0}}\lambda_{i}$ equals
\begin{eqnarray}
M_{X}(s)&\!\!\!=\!\!\!&\mathbb{E}\left[\e^{sX}\right]\\
&\!\!\!=\!\!\!&\frac{1}{c}\int_{\mathcal{D}}\det\left(\lambda_{i}^{j-1}\right)\det\left(G_{0,n}^{n,0}\left(\lambda_{i}\left|\begin{array}{c}\{\},\{\}\\\{\nu_{n},\ldots,\nu_{2},\nu_{1}+j-1\},\{\}\\\end{array}\right.\right)\right)\prod_{i=1}^{K_{0}}\e^{s\lambda_{i}}\mathrm{d}\mathbf{\Lambda}\nonumber\\
&\!\!\!=\!\!\!&\frac{1}{c}\det\left(\int_{0}^{\infty}\e^{s\lambda}\lambda^{j-1}G_{0,n}^{n,0}\left(\lambda\left|\begin{array}{c}\{\},\{\}\\\{\nu_{n},\ldots,\nu_{2},\nu_{1}+i-1\},\{\}\\\end{array}\right.\right)\mathrm{d}\lambda\right),\label{eq:MGFi}
\end{eqnarray}
where $\mathcal{D}=\{0\leq \lambda_{K_{0}}\leq\ldots\leq \lambda_{1}<\infty\}$ and the last equality is established by the Andr\'{e}ief integral~(\ref{eq:AI}). Using the shifting property~\cite[Eq.~9.315]{2007GR} of Meijer's G-function, we have
\begin{eqnarray}\label{eq:MSP}
\lambda^{j-1}G_{0,n}^{n,0}\left(\lambda\left|\begin{array}{c}\{\},\{\}\\\{\nu_{n},\ldots,\nu_{2},\nu_{1}+i-1\},\{\}\\\end{array}\right.\right)=\nonumber\\
G_{0,n}^{n,0}\left(\lambda\left|\begin{array}{c}\{\},\{\}\\\{\nu_{n}+j-1,\ldots,\nu_{2}+j-1,\nu_{1}+i+j-2\},\{\}\\\end{array}\right.\right).
\end{eqnarray}
Inserting~(\ref{eq:MSP}) and the identity
\begin{equation}
\e^{s\lambda}=G_{0,1}^{1,0}\left(-s\lambda\left|\begin{array}{c}\{\},\{\}\\\{0\},\{\}\\\end{array}\right.\right)
\end{equation}
into~(\ref{eq:MGFi}), the remaining integral is calculated by using~\cite[Eq.~7.811]{2007GR} as
\begin{eqnarray}
&&\int_{0}^{\infty}G_{0,1}^{1,0}\left(-s\lambda\left|\begin{array}{c}\{\},\{\}\\\{0\},\{\}\\\end{array}\right.\right)G_{0,n}^{n,0}\left(\lambda\left|\begin{array}{c}\{\},\{\}\\\{\nu_{n}+j-1,\ldots,\nu_{2}+j-1,\nu_{1}+i+j-2\},\{\}\\\end{array}\right.\right)\mathrm{d}\lambda\nonumber\\
&=&G_{1,n}^{n,1}\left(-\frac{1}{s}\left|\begin{array}{l}\{1\},\{\}\\\{\nu_{n}+j,\ldots,\nu_{2}+j,\nu_{1}+i+j-1\},\{\}\\\end{array}\right.\right).
\end{eqnarray}
So far we have established that
\begin{equation}\label{eq:MeijerMGF}
M_{X}(s)=\frac{1}{c}\det\left(G_{1,n}^{n,1}\left(-\frac{1}{s}\left|\begin{array}{l}\{1\},\{\}\\\{\nu_{n}+j,\ldots,\nu_{2}+j,\nu_{1}+i+j-1\},\{\}\\\end{array}\right.\right)\right).
\end{equation}
The Meijer's G-function in the above determinant is of the type that can be reduced to certain hypergeometric function~\cite[Eq.~9.348]{2007GR},
which by definition admits a formal power series expansion~\cite{1969Niven}. Namely, we have
\begin{eqnarray}
&&G_{1,n}^{n,1}\left(-\frac{1}{s}\left|\begin{array}{l}\{1\},\{\}\\\{\nu_{n}+j,\ldots,\nu_{2}+j,\nu_{1}+i+j-1\},\{\}\\\end{array}\right.\right)\\
&=&\left(\prod_{q=2}^{n}\Gamma(j+v_{q})\right)\sum_{t=0}^{\infty}\Gamma(i+j+\nu_{1}+t-1)\left(\prod_{q=2}^{n}(j+\nu_{q})_{t}\right)s^{t}/t!,\nonumber
\end{eqnarray}
where $(j+v_{q})_{t}$ is the Pochhammer symbol~(\ref{eq:Poch}). Inserting the above series into~(\ref{eq:MeijerMGF}) completes the proof of Proposition~\ref{p:MGF}.
\end{IEEEproof}

\section{Proof of Proposition~\ref{p:ExMo}}\label{a:ExMo}
\begin{IEEEproof}
By the definition of MGF~(\ref{eq:defMGF}), the $m$-th moment of $X=\sum_{i=1}^{K_{0}}\lambda_{i}$ equals the coefficient of $s^{m}$ in~(\ref{eq:MGF}) multiplied by $m!$. For $\mathbb{E}\left[X^{m}\right]$ it is sufficient to calculate the coefficient of $s^{m}$ from the determinant
\begin{equation}\label{eq:mdet}
\det\left(\sum_{t=0}^{m}\Gamma(i+j+\nu_{1}+t-1)\left(\prod_{q=2}^{n}(j+\nu_{q})_{t}\right)\frac{s^{t}}{t!}\right),
\end{equation}
whose entries are truncated series of~(\ref{eq:MGF}). By the multi-linearity property of matrix determinant, the above determinant can be written as a sum of $(m+1)^{K_{0}}$ determinants. Out of these, the sum of the determinants of the form
\begin{equation}
\sum_{a_{1}+\cdots+a_{K_{0}}=m}\det\left(\frac{\Gamma(i+j+a_{j}+\nu_{1}-1)\prod_{q=2}^{n}(j+\nu_{q})_{a_{j}}}{a_{j}!}\right)
\end{equation}
is coefficient of $s^{m}$, where $a_{i}\in\{0,\dots,m\}$ for $i=1,\ldots,K_{0}$. As a result, the $m$-th moment of $X$ equals
\begin{equation}\label{eq:IpEM}
\mathbb{E}\left[X^{m}\right]=\frac{m!}{\prod_{j=1}^{K_{0}}\Gamma(j)\Gamma(j+\nu_{1})}\sum_{L}\det\big(\Gamma(i+j+a_{j}+\nu_{1}-1)\big)\prod_{j=1}^{K_{0}}\prod_{i=2}^{n}\frac{\Gamma(j+a_{j}+\nu_{i})}{\Gamma(a_{j}+1)\Gamma(j+\nu_{i})},
\end{equation}
where $L$ denotes $a_{1}+\cdots+a_{K}=m$. The determinant in~(\ref{eq:IpEM}) can be simplified as
\begin{eqnarray}
\det\big(\Gamma(i+j+a_{j}+\nu_{1}-1)\big)&=&\det\big((a_{j}+j)_{i+\nu_{1}-1}\big)\prod_{j=1}^{K_{0}}\Gamma(a_{j}+j)\label{eq:I2pEM}\\
&=&\prod_{j=1}^{K_{0}}(a_{j}+j)_{\nu_{1}}\prod_{1\leq i<j \leq K_{0}}\left(a_{i}-a_{j}+i-j\right)\prod_{j=1}^{K_{0}}\Gamma(a_{j}+j),\label{eq:I3pEM}
\end{eqnarray}
where the determinant $\det\big((a_{j}+j)_{i+\nu_{1}-1}\big)$ in~(\ref{eq:I2pEM}) is identified to be a Vandermonde determinant~(\ref{eq:Van}) by
first factoring out the term $(a_{j}+j)_{\nu_{1}}$ from each column and then extracting from the remaining determinant the $i$-th row a suitable linear combination of previous $i-1$ rows. Inserting~(\ref{eq:I3pEM}) into~(\ref{eq:IpEM}) completed the proof of Proposition~\ref{p:ExMo}.
\end{IEEEproof}

\section{Proof of Lemma~\ref{l:Det}}\label{a:Det}
\begin{IEEEproof}
By the property of matrix determinant, the determinant in~(\ref{eq:DetL2}) remains unchanged by replacing the $i$-th row $r_{i}$ with $r_{i}-(i-1+\nu_{1})r_{i-1}$ for $i=2,\ldots,K_{0}$, namely
\begin{eqnarray*}
&&\det\left(\begin{array}{cccc}
\Gamma(1+\nu_{1}) & \Gamma(2+\nu_{1}) & \cdots & \Gamma(K_{0}+\nu_{1}+m) \\
\Gamma(2+\nu_{1}) & \Gamma(3+\nu_{1}) & \cdots & \Gamma(K_{0}+1+\nu_{1}+m) \\
\vdots & \vdots & \ddots & \vdots \\
\Gamma(K_{0}+\nu_{1}) & \Gamma(K_{0}+1+\nu_{1}) & \cdots & \Gamma(2K_{0}-1+\nu_{1}+m) \\
\end{array}\right)\\
&=&(K_{0}-2)!(K_{0}-1+m)\det\left(\begin{array}{cccc}
\Gamma(1+\nu_{1}) & \Gamma(2+\nu_{1}) & \cdots & \Gamma(K_{0}+\nu_{1}+m) \\
0 & \Gamma(2+\nu_{1}) & \cdots & \Gamma(K_{0}+\nu_{1}+m) \\
0 & \Gamma(3+\nu_{1}) & \cdots & \Gamma(K_{0}+1+\nu_{1}+m) \\
\vdots & \vdots & \ddots & \vdots \\
0 & \Gamma(K_{0}+\nu_{1}) & \cdots & \Gamma(2K_{0}-2+\nu_{1}+m) \\
\end{array}\right).\nonumber
\end{eqnarray*}
Repeating the above procedure $K_{0}-2$ times, we arrive at
\begin{eqnarray*}
&&\det\left(\begin{array}{cccc}
\Gamma(1+\nu_{1}) & \Gamma(2+\nu_{1}) & \cdots & \Gamma(K_{0}+\nu_{1}+m) \\
\Gamma(2+\nu_{1}) & \Gamma(3+\nu_{1}) & \cdots & \Gamma(K_{0}+1+\nu_{1}+m) \\
\vdots & \vdots & \ddots & \vdots \\
\Gamma(K_{0}+\nu_{1}) & \Gamma(K_{0}+1+\nu_{1}) & \cdots & \Gamma(2K_{0}-1+\nu_{1}+m) \\
\end{array}\right)\\
&=&\prod_{i=1}^{K_{0}-1}(K_{0}-i-1)!(K_{0}-i+m)\det\left(\begin{array}{ccccc}
\Gamma(1+\nu_{1}) & \Gamma(2+\nu_{1}) & \Gamma(3+\nu_{1}) & \cdots & \Gamma(K_{0}+\nu_{1}+m) \\
0 & \Gamma(2+\nu_{1}) & \Gamma(3+\nu_{1}) & \cdots & \Gamma(K_{0}+\nu_{1}+m) \\
0 & 0 & \Gamma(3+\nu_{1}) & \cdots & \Gamma(K_{0}+\nu_{1}+m) \\
\vdots & \vdots & \vdots & \ddots & \vdots \\
0 & 0 & 0 & \cdots & \Gamma(K_{0}+\nu_{1}+m)\\
\end{array}\right)\nonumber\\
&=&\prod_{i=1}^{K_{0}-1}(K_{0}-i-1)!(K_{0}-i+m)\left(\prod_{i=1}^{K_{0}-1}\Gamma(i+\nu_{1})\right)\Gamma(K_{0}+\nu_{1}+m)\\
&=&\frac{\Gamma(m+\nu_{1}+K_{0})\Gamma(m+K_{0})}{\Gamma(m+1)}\prod_{i=1}^{K_{0}-1}\Gamma(i)\Gamma(i+\nu_{1}),
\end{eqnarray*}
where in the last equality we utilized the definition~(\ref{eq:Poch}). This completes the proof.
\end{IEEEproof}

\section{Proof of Corollary~\ref{c:mo}}\label{a:mo}
\begin{IEEEproof}
For the first moment $\mathbb{E}\left[X\right]$, one needs to extract the coefficient of $s$ from the determinant
\begin{equation}\label{eq:m1det}
\det\left(\Gamma(i+j+\nu_{1}-1)+\Gamma(i+j+\nu_{1})\left(\prod_{q=2}^{n}(j+\nu_{q})\right)s\right).
\end{equation}
Since the every entry in~(\ref{eq:m1det}) consists of a sum of two terms, by the multi-linearity property of matrix determinant, one can write~(\ref{eq:m1det}) as a sum of $2^{K_{0}}$ determinants. Out of the $2^{K_{0}}$ determinants, the only non-zero contribution to the coefficient of $s$ is given by
\begin{equation}\label{eq:DETm1}
\begin{array}{c:c}
\det\Bigg(\Gamma\big(i+j+\nu_{1}-1\big)&\Gamma\big(i+K_{0}+\nu_{1}\big)\prod_{q=2}^{n}(K_{0}+\nu_{q})\Bigg)
\end{array}
\end{equation}
\begin{equation*}
\qquad=\left(\prod_{i=2}^{n}(K_{0}+\nu_{i})\right)\Gamma(\nu_{1}+K_{0}+1)\Gamma(K_{0}+1)\prod_{j=2}^{K_{0}-1}\Gamma(j)\Gamma(j+\nu_{1}),
\end{equation*}
where the equality is obtained by Lemma~\ref{l:Det} for $m=1$. The determinant~(\ref{eq:DETm1}) corresponds to the case that the only non-zero $\{a_{i}\}_{i=1}^{K_{0}}$ is $\{a_{K_{0}}=1\}$ in~(\ref{eq:ExMo}). Finally, taking into account the normalization, we have
\begin{eqnarray}
\mathbb{E}\left[X\right]&=&\frac{\big(\prod_{i=2}^{n}(K_{0}+\nu_{i})\big)\Gamma(\nu_{1}+K_{0}+1)\Gamma(K_{0}+1)\prod_{j=2}^{K_{0}-1}\Gamma(j)\Gamma(j+\nu_{1})}{\prod_{j=1}^{K_{0}}\Gamma(j)\Gamma(j+\nu_{1})}\nonumber\\
&=&\prod_{i=0}^{n}(K_{0}+\nu_{i})=\prod_{i=0}^{n}K_{i}.
\end{eqnarray}
This completes the proof of~(\ref{eq:m1}).

For the second moment $\mathbb{E}\left[X^{2}\right]$, we need to calculate the coefficient of $s^{2}$ from the determinant
\begin{eqnarray}\label{eq:m2det}
&&\det\Bigg(\Gamma(i+j+\nu_{1}-1)+\Gamma(i+j+\nu_{1})\left(\prod_{q=2}^{n}(j+\nu_{q})\right)s\nonumber\\
&&+\Gamma(i+j+\nu_{1}+1)\left(\prod_{q=2}^{n}(j+\nu_{q})_{2}\right)\frac{s^{2}}{2}\Bigg).
\end{eqnarray}
By using the multi-linearity property, the determinant~(\ref{eq:m2det}) can be written as a sum of $3^{K_{0}}$ determinants. Among these, the non-zero contribution to the coefficient of $s^{2}$ is the sum of the following three determinants
\begin{equation}\label{eq:m2Det1}
\begin{array}{c:c}
\det\Bigg(\Gamma\big(i+j+\nu_{1}-1\big)&\displaystyle\frac{\Gamma(i+K_{0}+\nu_{1}+1)\prod_{q=2}^{n}(K_{0}+\nu_{q})_{2}}{2}\Bigg),
\end{array}
\end{equation}
\begin{equation*}
\begin{array}{c:c:c}
\det\Bigg(\Gamma\big(i+j+\nu_{1}-1\big)&\displaystyle\Gamma(i+K_{0}+\nu_{1}-1)\prod_{q=2}^{n}(K_{0}+\nu_{q}-1)&
\end{array}
\end{equation*}
\begin{equation}\label{eq:m2Det2}
\Gamma(i+K_{0}+\nu_{1})\prod_{q=2}^{n}(K_{0}+\nu_{q})\Bigg),
\end{equation}
\begin{equation*}
\begin{array}{c:c:c}
\det\Bigg(\Gamma\big(i+j+\nu_{1}-1\big)&\displaystyle\frac{\Gamma(i+K_{0}+\nu_{1})\prod_{q=2}^{n}(K_{0}+\nu_{q}-1)_{2}}{2}&
\end{array}
\end{equation*}
\begin{equation}\label{eq:m2Det3}
\Gamma(i+K_{0}+\nu_{1}-1)\Bigg),
\end{equation}
where $i=1,\ldots,K_{0}$, $j=1,\ldots,K_{0}-1$ in~(\ref{eq:m2Det1}) and $i=1,\ldots,K_{0}$, $j=1,\ldots,K_{0}-2$ in~(\ref{eq:m2Det2}) and~(\ref{eq:m2Det3}). The determinants~(\ref{eq:m2Det1}), (\ref{eq:m2Det2}) and~(\ref{eq:m2Det3}) correspond to the cases that the non-zero $\{a_{i}\}_{i=1}^{K_{0}}$ are $\{a_{K_{0}}=2\}$, $\{a_{K_{0}-1}=a_{K_{0}}=1\}$ and $\{a_{K_{0}-1}=2\}$ in~(\ref{eq:ExMo}), respectively. The sum of the three determinants can be simplified to
\begin{equation}\label{eq:m2co}
\frac{\prod_{i=2}^{n}(K_{0}+\nu_{i})_{2}}{2}d_{1}+\frac{\prod_{i=2}^{n}(K_{0}+\nu_{i}-1)_{2}}{2}d_{2},
\end{equation}
where the determinants
\begin{equation}
d_{1}=\begin{array}{c:c}
\det\Bigg(\Gamma\big(i+j+\nu_{1}-1\big)&\displaystyle\Gamma(i+K_{0}+\nu_{1}+1)\Bigg),
\end{array}
\end{equation}
and
\begin{equation}
d_{2}=\begin{array}{c:c:c}
\det\Bigg(\Gamma\big(i+j+\nu_{1}-1\big)&\Gamma\big(i+K_{0}+\nu_{1}-1\big)&\Gamma\big(i+K_{0}+\nu_{1}\big)\Bigg).
\end{array}
\end{equation}
Using Lemma~\ref{l:Det} with $m=2$, we have $d_{1}=\frac{\Gamma(K_{0}+\nu_{1}+2)\Gamma(K_{0}+2)}{2}\prod_{j=1}^{K_{0}-1}\Gamma(j)\Gamma(j+\nu_{1})$. The determinant $d_{2}$ can be obtained by the known result of the moments of $X$ for $n=1$
\begin{equation}\label{eq:mon1}
\mathbb{E}\left[X^{m}\right]=\left(K_{0}K_{1}\right)_{m}.
\end{equation}
Namely, putting $n=1$ in~(\ref{eq:m2co}), we have
\begin{equation}
\frac{2!}{\prod_{j=1}^{K_{0}}\Gamma(j)\Gamma(j+\nu_{1})}\left(\frac{d_{1}}{2}+\frac{d_{2}}{2}\right)=K_{0}K_{1}(K_{0}K_{1}+1),
\end{equation}
from which $d_{2}$ is solved as $d_{2}=\frac{K_{0}K_{1}(K_{0}-1)(K_{1}-1)}{2}\prod_{j=1}^{K_{0}}\Gamma(j)\Gamma(j+\nu_{1})$. With the expressions $d_{1}$ and $d_{2}$, we finally have
\begin{eqnarray}
\mathbb{E}\left[X^{2}\right]&=&\frac{2!}{\prod_{j=1}^{K_{0}}\Gamma(j)\Gamma(j+\nu_{1})}\left(\frac{\prod_{i=2}^{n}(K_{0}+\nu_{i})_{2}}{2}d_{1}+\frac{\prod_{i=2}^{n}(K_{0}+\nu_{i}-1)_{2}}{2}d_{2}\right)\nonumber\\
&=&\frac{\prod_{i=0}^{n}K_{i}}{2}\left(\prod_{i=0}^{n}(K_{i}+1)+\prod_{i=0}^{n}(K_{i}-1)\right).
\end{eqnarray}
This completes the proof of~(\ref{eq:m2}). By following the same principle of identifying relevant determinants, and after some tedious but straightforward calculations, we arrive at~(\ref{eq:m3}). The excessively lengthy derivation is, however, omitted here.
\end{IEEEproof}

\section{Proof of Corollary~\ref{c:amo}}\label{a:amo}
\begin{IEEEproof}
It is seen that among the $(m+1)^{K_{0}}$ determinants in the multi-linearity expansion of~(\ref{eq:mdet}) the following term
\begin{equation}\label{eq:anymDet}
\begin{array}{c:c}
\det\Bigg(\Gamma\big(i+j+\nu_{1}-1\big)&\displaystyle\frac{\Gamma\big(i+K_{0}+\nu_{1}+m-1\big)\prod_{q=2}^{n}(K_{q})_{m}}{m!}\Bigg)
\end{array}
\end{equation}
gives the largest contribution to the coefficient of $s^{m}$ for large $n$. This determinant is formed by selecting the constant terms from each of the first $K_{0}-1$ columns to form the corresponding first $K_{0}-1$ columns in~(\ref{eq:anymDet}), whereas the last column is formed by selecting the $s^{m}$ terms from the last column. As a result, the largest $n$-dependent term $\prod_{i=2}^{n}(K_{i})_{m}$ has been found. The determinant~(\ref{eq:anymDet}) corresponds to the case that the only non-zero $\{a_{i}\}_{i=1}^{K_{0}}$ is $\{a_{K_{0}}=m\}$ in~(\ref{eq:ExMo}). The corresponding $m$-th moment of $X$ can now be approximated by the dominant term~(\ref{eq:anymDet}) as
\begin{eqnarray}
\mathbb{E}\left[X^{m}\right]&\approx&\frac{m!}{\prod_{j=1}^{K_{0}}\Gamma(j)\Gamma(j+\nu_{1})}\frac{\prod_{i=2}^{n}(K_{i})_{m}}{m!}\begin{array}{c:c}
\det\Bigg(\Gamma\big(i+j+\nu_{1}-1\big)&\displaystyle\Gamma\big(i+K_{0}+\nu_{1}+m-1\big)\Bigg)
\end{array}\nonumber\\
&=&\frac{\prod_{i=2}^{n}(K_{i})_{m}}{\prod_{j=1}^{K_{0}}\Gamma(j)\Gamma(j+\nu_{1})}\frac{\Gamma(m+\nu_{1}+K_{0})\Gamma(m+K_{0})}{\Gamma(m+1)}\prod_{j=1}^{K_{0}-1}\Gamma(j)\Gamma(j+\nu_{1})\\
&=&\frac{\prod_{i=0}^{n}\left(K_{i}\right)_{m}}{m!},\label{eq:LTn}
\end{eqnarray}
where we have invoked Lemma~\ref{l:Det}. This completes the proof of~(\ref{eq:mApp}).

Similarly to the construction of~(\ref{eq:anymDet}), the next leading order terms to the coefficient of $s^{m}$ for large $n$ are found to be
\begin{equation*}
\begin{array}{c:c:c}
\det\Bigg(\Gamma\big(i+j+\nu_{1}-1\big)&\displaystyle\frac{\Gamma(i+K_{0}+\nu_{1}+m-2)\prod_{q=2}^{n}(K_{0}+\nu_{q}-1)_{m}}{m!}&
\end{array}
\end{equation*}
\begin{equation}
\Gamma\big(i+K_{0}+\nu_{1}-1\big)\Bigg),
\end{equation}
and
\begin{equation*}
\begin{array}{c:c:c}
\det\Bigg(\Gamma\big(i+j+\nu_{1}-1\big)&\displaystyle\Gamma\big(i+K_{0}+\nu_{1}-1\big)\prod_{q=2}^{n}(K_{0}+\nu_{q}-1)&
\end{array}
\end{equation*}
\begin{equation}
\frac{\Gamma\big(i+K_{0}+\nu_{1}+m-2\big)\prod_{q=2}^{n}(K_{0}+\nu_{q})_{m-1}}{(m-1)!}\Bigg).
\end{equation}
Thus, the resulting second leading order term of $\mathbb{E}\left[X^{m}\right]$ can be written as
\begin{equation}
\varphi\prod_{i=2}^{n}(K_{i}-1)_{m},
\end{equation}
where the constant $\varphi$ denotes the factors that do not depend on $n$. In the limit of large $n$, the ratio of the second leading order term to the first order term~(\ref{eq:LTn}) is computed as
\begin{eqnarray}
\lim_{n\to\infty}\frac{\varphi\prod_{i=2}^{n}(K_{i}-1)_{m}}{\prod_{i=0}^{n}\left(K_{i}\right)_{m}/m!}&=&\frac{m!\varphi}{(K_{0})_{m}(K_{1})_{m}}\lim_{n\to\infty}\prod_{i=2}^{n}\frac{(K_{i}-1)_{m}}{(K_{i})_{m}}\\
&=&\frac{m!\varphi}{(K_{0})_{m}(K_{1})_{m}}\lim_{n\to\infty}\prod_{i=2}^{n}\frac{K_{i}-1}{K_{i}+m-1}=0.
\end{eqnarray}
Bearing in mind that all the other lower order terms vanish in the same fashion completes the proof of~(\ref{eq:conApp}).
\end{IEEEproof}

\section{Proof of Proposition~\ref{p:conv}}\label{a:conv}
\begin{IEEEproof}
As the joint density of the non-zero eigenvalues of $\mathbf{HH}^{\dag}$ is invariant under the permutations of the matrix dimensions~\cite{2013AIK,2013IK}, without loss of generality we set $K_{1}=K'$, i.e. $\rho_{1}=1$. Define $\mathbf{H}'=\mathbf{H}_{n-1}\cdots\mathbf{H}_{2}/\sqrt{\prod_{i=2}^{n-1}K_i}$ and perform singular value decomposition $\mathbf{H}'=\mathbf{U}\mathbf{\Sigma}\mathbf{V}^\dag$, where $\mathbf{U}$, $\mathbf{V}$ are unitary matrices and $\left(\mathbf{\Sigma}\right)_{i,i}=s_{i}$, $i=1,\dots,K_{1}$, denote the corresponding singular values. As a result, (\ref{ch}) can be rewritten as
\begin{equation}
\mathbf{H}=\frac{\mathbf{H}_{n}\mathbf{U}\mathbf{\Sigma}\mathbf{V}^\dag\mathbf{H}_1}{\sqrt{K_1}},
\end{equation}
which has the same distribution as
\begin{equation}
\frac{\mathbf{H}_{n}\mathbf{\Sigma}\mathbf{H}_1}{\sqrt{K_1}}=\frac{1}{\sqrt{K_1}}\sum_{i=1}^{K_1}s_{i}\mathbf{h}_{i}\mathbf{g}_{i},
\end{equation}
with $\mathbf{h}_i$ being the $i$-th column of $\mathbf{H}_n$ and $\mathbf{g}_i$ being the $i$-th row of $\mathbf{H}_1$. Now we have
\begin{equation}
vec(\mathbf{H})=\sum_{i=1}^{K_{1}}\mathbf{z}_{i}=\frac{1}{\sqrt{K_1}}\sum_{i=1}^{K_1}s_{i}\,vec\left(\mathbf{h}_{i}\mathbf{g}_{i}\right),
\end{equation}
where conditioned on $\mathbf{H}'$, $\mathbf{z}_i$ are mutually independent circular symmetric random vectors as $\mathbf{h}_i$ and $\mathbf{g}_i$ are mutually independent. Under this setting, both a) and b) hold~\cite[Coroll.~7]{2011Levin} if
\begin{equation}\label{lim}
\lim_{K_{1}\rightarrow\infty}\mathbb{E}\left\|\mathbf{C}^{-1/2}\mathbf{z}_{i}\right\|_{2}^{3}=0,
\end{equation}
where $\left\|\cdot\right\|_{2}$ denotes the vector Euclidean norm and
\begin{equation} \mathbf{C}=\mathbb{E}\left[vec(\mathbf{H})vec(\mathbf{H})^\dag\right]=\frac{\mathbf{I}_{K_{0}K_{n}}}{K_1}\sum_{i=1}^{K_1}s_{i}^2,
\end{equation}
is the covariance matrix of $vec(\mathbf{H})$. The rest of proof is devoted to show~(\ref{lim}). Specifically,
\begin{eqnarray}
\sum_{i=1}^{K_1}\mathbb{E}\left\|\mathbf{C}^{-1/2}\mathbf{z}_i\right\|_{2}^3&\le&\sum_{i=1}^{K_1}\left(\mathbb{E}\left\|\mathbf{C}^{-1/2}\mathbf{z}_i\right\|_{2}^4\right)^{3/4}\label{eq3Moa}\\
&\le&\left\|\mathbf{C}^{-1/2}\right\|_{\text{F}}^{3}\sum_{i=1}^{K_1}\left(\mathbb{E}\left\|\mathbf{z}_i\right\|_{2}^4\right)^{3/4}\label{eq3Mob}\\
&=&\frac{1}{K_1^{1/2}}\left\|\mathbf{C}^{-1/2}\right\|_{\text{F}}^{3}\left(\mathbb{E}\|\mathbf{h}_1\|_{2}^{4}\mathbb{E}\|\mathbf{g}_1\|_{2}^4\right)^{3/4}\frac{1}{K_{1}}\sum_{i=1}^{K_1} s_i^3\label{eq3Moc},
\end{eqnarray}
where~(\ref{eq3Moa}) is due to Lyapunov~\cite[Th.~3.4.1]{1963Fisz} and~(\ref{eq3Mob}) is obtained by Cauchy-Schwarz inequality. As $K_{1}$ goes to infinity, $\left\|\mathbf{C}^{-1/2}\right\|_{\text{F}}^{3}$ is finite since
\begin{equation}
\lim_{K_1\rightarrow\infty}\mathbf{C}=\mathbf{I}_{K_{0}K_{n}} \lim_{K_1\rightarrow\infty}\frac{\mathrm{tr}\left(\mathbf{H}'(\mathbf{H}')^\dag\right)}{K_1}=\mathbf{I}_{K_{0}K_{n}},\label{eqC}
\end{equation}
and $\mathbb{E}\|\mathbf{h}_1\|_{2}^{4}\mathbb{E}\|\mathbf{g}_1\|_{2}^4$ is finite as $\mathbf{h}_1$ and $\mathbf{g}_1$ are finite dimensional Gaussian vectors. Moreover,
\begin{eqnarray}
\lim_{K_{1}\rightarrow\infty}\frac{1}{K_{1}}\sum_{i=1}^{K_1}s_{i}^{3}&=&\int x^{3/2}\,\mathrm{d}F_{\lambda'}(x)\label{b1}\\
&\leq&\left(\int x^{2}\,\mathrm{d}F_{\lambda'}(x)\right)^{3/4}=\left(\mathbb{E}[(\lambda')^2]\right)^{3/4}\label{b},
\end{eqnarray}
where $F_{\lambda'}(x)$ denotes the limiting distribution of an arbitrary eigenvalue of $\mathbf{H}'(\mathbf{H}')^\dag$ and the inequality in~(\ref{b}) is established by~\cite[Th.~3.4.1]{1963Fisz}. Here, $\mathbb{E}\left[(\lambda')^2\right]$ exists and is finite~\cite[Th.~3]{2002Muller}. Thus, as $K_1$ goes to infinity, $\mathbb{E}\left\|\mathbf{C}^{-1/2}\mathbf{z}_{i}\right\|_{2}^{3}$ approaches zero with at least the same rate as $(K')^{-1/2}$ approaches zero. This completes the proof.
\end{IEEEproof}

\section*{Acknowledgment}
L. Wei is supported by the Finnish Centre of Excellence in Computational Inference Research, Academy of Finland (Grant 251170). Z. Zheng is supported by Academy of Finland (Grant 254299). J. Corander is supported by European Research Council (Grant 239784). G. Taricco is supported by NEWCOM\#, WP 1.1.1.

\end{document}